\documentclass[twocolumn,10pt]{article}

\usepackage[a4paper,margin=1in]{geometry}

\usepackage[T1]{fontenc}
\usepackage[utf8]{inputenc}
\usepackage{mathptmx}      
\usepackage{microtype}

\usepackage{amsmath}
\usepackage{amssymb}

\usepackage{siunitx}
\usepackage{graphicx}
\graphicspath{{figures/}}
\usepackage{booktabs}
\usepackage[font=small,labelfont=bf,labelsep=period]{caption}

\usepackage{stfloats}          
\usepackage{authblk}

\usepackage{url}
\usepackage[colorlinks=true,linkcolor=blue,citecolor=blue,urlcolor=blue]{hyperref}

\usepackage{setspace}
\title{\bfseries Numerical Investigation of a 3D End-Firing Antenna
Array Based on Two-Photon Polymerization on Thin-Film Lithium Niobate
for Optical Beam Steering}

\author[1]{David Trop}
\author[1,*]{Boris Desiatov}
\affil[1]{Faculty of Engineering, Bar-Ilan University, Ramat Gan 5290002, Israel}
\affil[*]{Corresponding author: \href{mailto:boris.desiatov@biu.ac.il}{boris.desiatov@biu.ac.il}}

\date{}

\begin{document}

\twocolumn[
  \begin{@twocolumnfalse}
  \maketitle
  \begin{abstract}
  \noindent
  Optical phased arrays (OPAs) are key components for solid-state beam
steering in emerging photonic technologies such as LiDAR, optical
communication, and adaptive optics. However, conventional integrated OPA
designs face trade-offs between bandwidth, steering range, and
fabrication complexity. Here we designed and numerically analyzed a novel
three-dimensional end-firing antenna array compatible with fabrication
using two-photon polymerization (2PP) directly on a thin-film lithium
niobate (TFLN) platform. By elevating the polymer antennas above the chip
surface, the design enables two-dimensional beam steering while
maintaining the broadband advantages of end-fire emission. Full-wave
electromagnetic simulations demonstrate transmission efficiencies up to
89.5\% over the \SIrange{1.4}{1.6}{\micro\meter} wavelength range,
achieving a field of view of \ang{24.9} $\times$ \ang{22.8} with
beamwidths of approximately \ang{1.7}. The architecture's compatibility
with electro-optic phase control and advanced array configurations
suggests significant potential for high-speed, low-loss beam steering
systems. This work establishes a foundation for scalable 3D photonic
phased arrays that bridge integrated optics with free-space beam
manipulation.
\end{abstract}
  \vspace{4pt}
  \vspace{1.2em}
  \end{@twocolumnfalse}
]

\section{Introduction}

Manipulating the wavefront of light beams is a fundamental capability
essential for modern optical technologies, underpinning advances in
applications ranging from high-resolution imaging~\cite{ref1,ref2} and
laser processing~\cite{ref3,ref4} to free-space optical
communications~\cite{ref5,ref6} and Light Detection and Ranging
(LiDAR)~\cite{ref7,ref8,ref9,ref10}. While traditional free-space beam
steering approaches---relying on components like spatial light modulators
(SLMs)~\cite{ref11,ref12} and adaptive mirrors~\cite{ref13,ref14}---offer
versatility, integrated photonic solutions are increasingly favored for
their promise of superior scalability, energy efficiency, and high-speed
operation in compact, solid-state systems~\cite{ref15,ref16,ref8}.

Optical phased arrays (OPAs) have emerged as the leading technology for
integrated, non-mechanical beam steering, directly translating the
well-established principles of radio frequency (RF) arrays to optical
wavelengths. However, the transition to the optical domain introduces
significant constraints. Achieving complete grating-lobe suppression
across the full steering range requires element spacings of half a
wavelength---a strict sub-micron condition at telecom wavelengths. For
the non-steered (broadside) case, element spacings condition of
approximately one wavelength is sufficient to push grating lobes to the
edge of the visible range, while also creating side-lobe suppression.
This fundamental spacing requirement drives intensive research into novel
integrated photonic platforms and antenna designs and has positioned
integrated photonic circuits (PICs) as ideal platforms for OPA
realization~\cite{ref17,ref18}.

Various material platforms have been explored for OPA implementation,
including silicon photonics (Si)~\cite{ref16,ref19}, silicon nitride
(Si$_3$N$_4$)~\cite{ref20}, and indium phosphide (InP)~\cite{ref21}.
While Si-based arrays are common, they typically rely on either the
thermo-optic effect, which provides low optical loss but is limited to
millisecond-scale response times, or the free-carrier plasma dispersion
effect, which achieves faster modulation at the expense of substantial
optical attenuation. These trade-offs have motivated the search for
alternative platforms that can simultaneously provide high-speed phase
modulation, low insertion loss, and broad operational bandwidth.

In this context, TFLN stands out as a highly attractive platform. Its
exceptionally strong electro-optic coefficient enables gigahertz-speed
phase modulation with minimal power consumption and low optical loss,
making it particularly well-suited for high-performance, rapid beam
steering applications~\cite{ref22,ref23}.

Integrated OPA designs traditionally employ either one or two-dimensional
(2D)~\cite{ref16,ref24} grating couplers or one-dimensional (1D) end-fire
antennas~\cite{ref25}. Grating coupler-based systems facilitate 2D
arrangements but are inherently wavelength-sensitive and require larger
element pitches, creating a critical trade-off between steering range and
beam quality~\cite{ref16}. End-fire antennas, conversely, allow for the
tight element spacing necessary for superior beam characteristics and
broadband operation but have historically been confined to 1D arrays due
to their planar, edge-emission geometry~\cite{ref25}. Realizing
high-performance, true 2D beam steering while maintaining the advantages
of end-fire emission remains a significant technological challenge.

\begin{figure*}[tp]
  \centering
  \includegraphics[width=\linewidth]{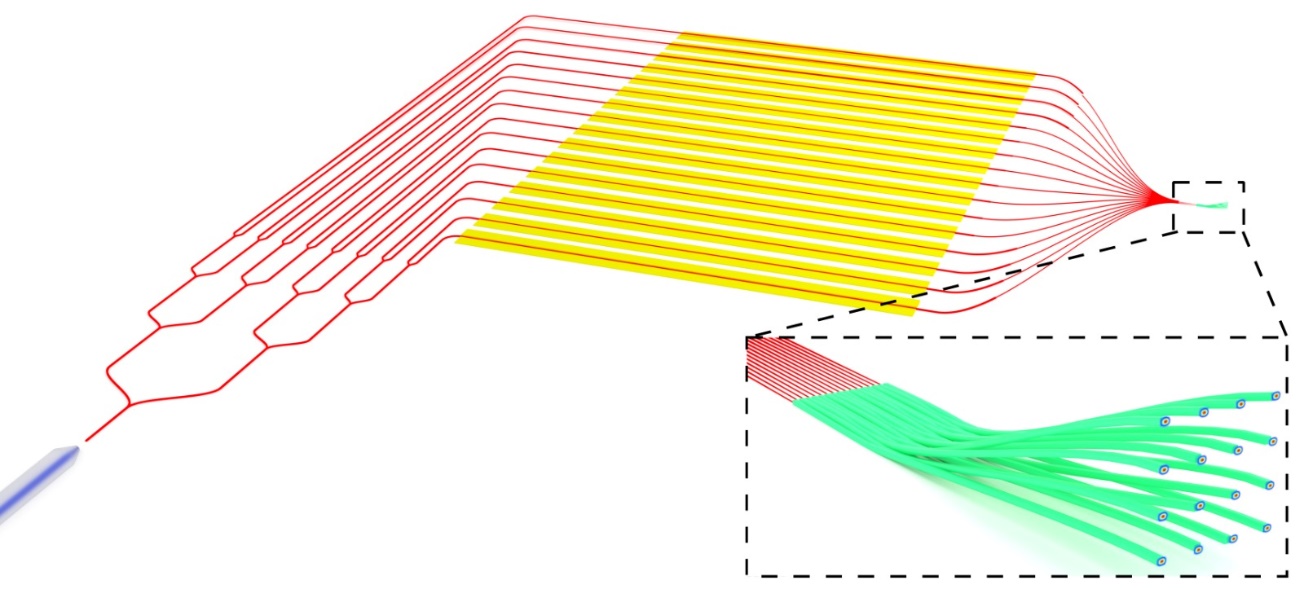}
  \caption{Schematic of the hybrid TFLN--polymer OPA. Light is coupled
  into a TFLN waveguide, distributed by Y-splitters, phase-modulated
  using electro-optic phase shifters, and emitted into free space through
  elevated three-dimensional polymer end-fire antennas. Inset shows a
  close-up of the 3D antenna geometry.}
  \label{fig:schematic}
\end{figure*}

In this work, we present a novel OPA architecture that integrates 3D
polymer waveguide antennas with the TFLN platform with the antenna
geometry realized through two-photon polymerization
(2PP)~\cite{ref26,ref27,ref28}. This approach successfully overcomes the
inherent 1D limitation of conventional end-fire arrays by elevating the
antenna structures above the TFLN chip surface, thereby enabling a dense
2D arrangement while preserving the broadband, high-efficiency
characteristics of end-fire emission. While previous works have
demonstrated 3D approaches to realize 2D end-fire antenna arrays by using
multilayer polymer photonic integrated circuits~\cite{ref29} and
ultrafast laser inscription in bulk glass~\cite{ref30}, the present work
introduces two key distinctions. First, we employ two-photon
polymerization (2PP) direct printing, a well-established and mature
fabrication technique in the photonic wire-bonding community, which
offers sub-\SI{200}{\nano\meter} feature resolution and fully flexible 3D
geometry control with straightforward integration on planar photonic
platforms. Second, and critically, by integrating this 3D antenna
architecture directly on the TFLN platform, we uniquely combine it for
the first time with the exceptional electro-optic properties of lithium
niobate, enabling gigahertz-speed phase modulation that is unavailable in
the silicon photonics or bulk glass platforms used in prior 3D OPA
demonstrations.

We conducted rigorous numerical simulations to optimize the antenna
geometry and mode transition, achieving a high transmission efficiency of
up to 89.5\% across the telecommunications band
(\SIrange{1.4}{1.6}{\micro\meter}). The resulting OPA demonstrates a
large field of view of \ang{24.9} $\times$ \ang{22.8} with an element
pitch of \SI{2}{\micro\meter}, maintaining narrow beam widths of
approximately \ang{1.7} $\times$ \ang{1.7}. These results validate the
elevated 3D end-fire architecture as a robust and broadband solution for
two-dimensional beam steering and highlight its compatibility with
advanced phased-array techniques.

\section{Device Design and Simulation Methods}

An optical phased array consists of several tightly coupled functional
building blocks, including waveguides, power distribution networks, phase
shifters, and optical antennas. Because the far-field performance of the
array depends on the coherent superposition of all emitting elements, the
optical efficiency, phase accuracy, and bandwidth of each individual
component directly impact the overall system performance. Consequently,
careful optimization of each subsystem is essential for achieving
low-loss, broadband, and high-fidelity beam steering.

This work studies the realization of an OPA, as illustrated in
Figure~\ref{fig:schematic}. First, light is launched into a single-mode
lithium niobate (LN) waveguide using an inverse taper. Next, the light is
split into several individual single-mode waveguides by a power
distribution system based on Y-splitter devices. The optical phase of
each individual channel is controlled by embedded electro-optical phase
shifters that utilize the Pockels effect, with individual voltages
applied to each EO phase shifter. Finally, the optical phased array image
is formed by outcoupling the optical signal into free space using
end-firing from an array of 3D optical antennas. These antennas are
fabricated in polymer on top of the LN waveguide to realize the 2D image
array.

The proposed OPA integrates this sequence of functional components into a
unified TFLN photonic platform combined with polymer-based 3D antennas
fabricated by two-photon polymerization (2PP). This hybrid integration
enables broadband, efficient, and scalable beam steering. The system
architecture focuses on transverse electric (TE) polarization at telecom
wavelengths (\SIrange{1.4}{1.6}{\micro\meter}), a spectral window widely
used in many OPA applications. While the design here is optimized for that
range, it can be reconfigured for other operational wavelengths as needed.

Full three-dimensional finite-difference time-domain (FDTD) simulations
were performed to evaluate light propagation, coupling efficiency, phase
control, and far-field beam formation. The following sections provide
detailed numerical analyses of the individual subsystems and summarize
their collective impact on overall device performance.

\subsection{Waveguide Platform Configuration}

In this work, we explored two TFLN platform configurations: a fully
etched structure with a \SI{300}{\nano\meter} lithium niobate layer and a
ridge waveguide comprising a \SI{100}{\nano\meter} slab with a
\SI{200}{\nano\meter} ridge. Both configurations incorporate a
\SI{1}{\micro\meter}-wide single-mode waveguide with an inverse taper
narrowing to \SI{250}{\nano\meter} for efficient coupling.

The ridge waveguide distributes the optical mode between the etched ridge
and the unetched slab, traditionally offering advantages such as reduced
sidewall scattering. However, for our specific application---coupling to
elevated polymer antenna structures---this distribution of optical energy
proves suboptimal. The fully etched configuration provides stronger
optical mode confinement within the LN core, with only evanescent-field
penetration into the silica substrate. Although a thicker LN layer could,
in principle, enhance confinement, practical fabrication challenges
associated with deeper etching necessitate a compromise at
\SI{300}{\nano\meter} thickness. This configuration demonstrates superior
performance in our application, particularly in terms of coupling
efficiency to the elevated polymer antennas.

\begin{figure*}[tp]
  \centering
  \includegraphics[width=\linewidth]{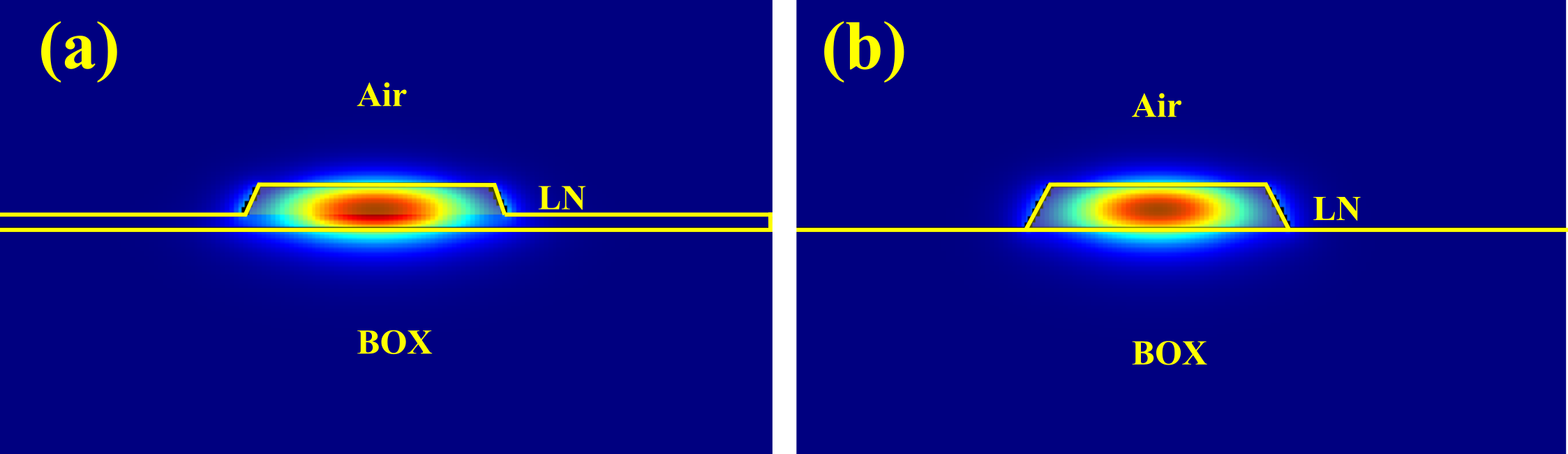}
  \caption{Simulated fundamental transverse-electric (TE) mode profiles
  of (a) fully etched and (b) ridge TFLN waveguides at a wavelength of
  \SI{1.55}{\micro\meter}. Stronger optical confinement in the fully
  etched structure enables more efficient coupling to elevated polymer
  antennas.}
  \label{fig:modes}
\end{figure*}

Figure~\ref{fig:modes} shows the simulated mode profiles of the
fundamental TE mode for both fully etched and partially etched waveguides
at a wavelength of \SI{1.55}{\micro\meter}. While there are similarities
in the optical power distribution between these structures, the device
performance is strongly influenced by the precise distribution of optical
energy within the waveguide. One key design parameter when selecting
specific waveguide geometries is the minimal bending radius, which
dictates the overall device footprint. For tightly confined TFLN
waveguides, we selected a minimum bending radius of \SI{50}{\micro\meter},
verified through numerical simulations to ensure low loss and compact
integration.

\subsection{Electro-Optic Phase Shifters}

Precise control over the optical phase of each individual channel is
essential for generating the desired electric field distribution in the
focal plane of the OPA. While the thermo-optic effect in lithium niobate
(LN) can be used for slow modulation applications, it imposes significant
constraints on device layout due to thermal crosstalk and heat
dissipation. In contrast, electro-optic (EO) phase shifters are not
limited by thermal coupling and can be densely integrated with minimal
interaction between neighboring elements.

\begin{figure*}[tp]
  \centering
  \includegraphics[width=\linewidth]{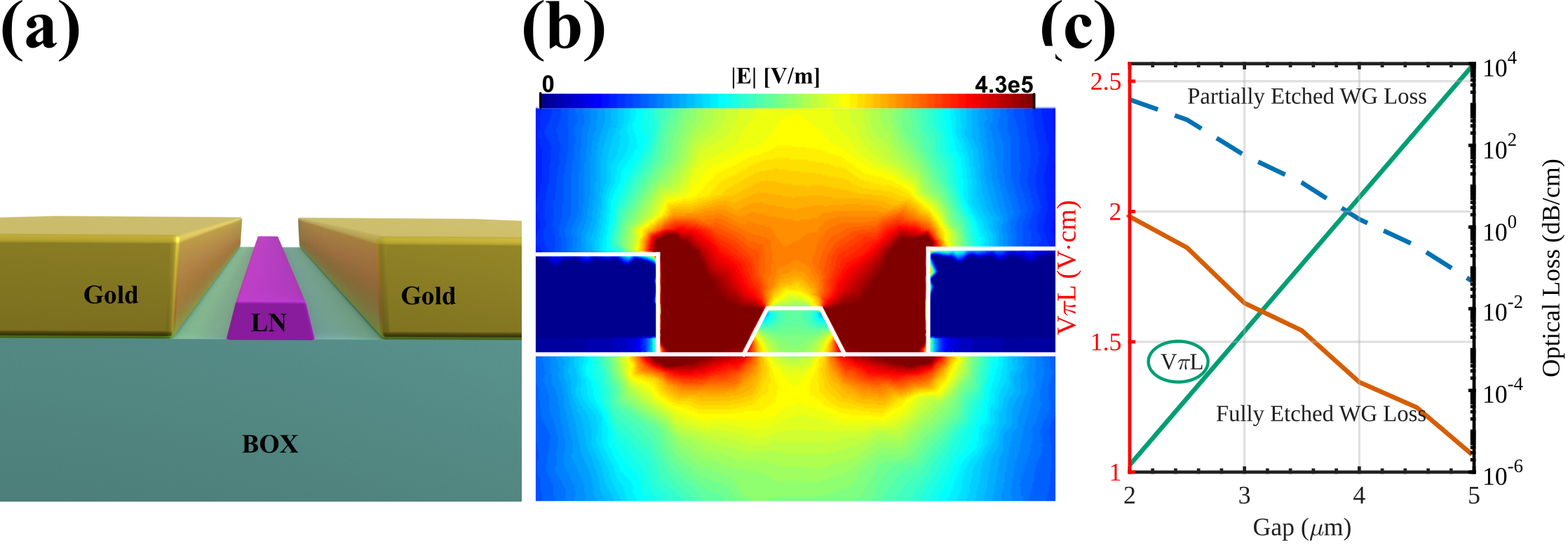}
  \caption{Electro-optic phase shifter design and performance. (a)
  Cross-sectional schematic of the electrode--waveguide configuration.
  (b) Simulated static electric field magnitude $|E|$ (V/m) distribution
  from Lumerical CHARGE solver at \SI{1.5}{\volt} applied voltage. (c)
  Calculated optical propagation loss and modulation efficiency as a
  function of electrode spacing.}
  \label{fig:phaseshifter}
\end{figure*}

Phase control in our system leverages TFLN's pronounced electro-optic
effect, particularly strong along the $z$-axis with an $r_{33}$
coefficient of approximately \SI{31}{\pico\meter\per\volt}. The induced
phase shift is given by
\begin{equation}
V_{\pi} = \frac{\lambda\, g}{n_{e}^{3}\, r_{33}\, L},
\end{equation}
where $\lambda$ represents the operating wavelength, $g$ denotes the gap
between the two electrodes beside the waveguide, $n_{e}$ is the
extraordinary refractive index of LN ($\approx 2.14$ at
\SI{1550}{\nano\meter}), and $L$ is the electrode length.
Figure~\ref{fig:phaseshifter} shows the schematic cross-section of the
phase shifter configuration.

A key metric for comparing EO phase shifter performance is the
$V_{\pi}{\cdot}L$ product, which quantifies intrinsic modulation
efficiency independent of device length. While reducing the electrode gap
$g$ theoretically enhances efficiency, practical designs must balance this
against increased optical losses due to metal absorption. A
\SIrange{4}{5}{\micro\meter} electrode gap provides an optimal
compromise, offering low propagation loss together with a moderate
operational voltage.

Recent demonstrations of TFLN-based optical phased arrays have achieved
switching speeds as fast as \SI{14.4}{\nano\second}, enabling beam
steering rates that far exceed those of thermo-optic
counterparts~\cite{ref31}. Our numerical analysis of metal-induced losses
and electro-optic field overlap confirms that the chosen gap
simultaneously satisfies the requirements for low optical attenuation and
efficient modulation. The precise electrode spacing may be further
adjusted to meet specific system constraints, such as device footprint or
maximum drive voltage.

The current design does not yet optimize the radio-frequency (RF)
characteristics of the electrodes. For low-speed applications, a simple
two-electrode configuration is sufficient and does not require matching
between optical and RF propagation velocities. However, for high-speed
operation exceeding \SI{1}{\giga\hertz}, impedance matching between the
electrodes and the optical waveguide becomes critical. This can be
achieved by tailoring the electrode geometry and oxide thickness to
equalize the propagation constants of both waves.

Table~\ref{tab:phaseshifters} summarizes the performance of recent phase
shifter technologies, including electro-optic, thermo-optic, and
carrier-injection implementations, highlighting the superior speed and
energy efficiency of EO phase shifters on the TFLN platform.

\begin{table*}[tp]
  \centering
  \caption{Performance comparison of phase shifters.}
  \label{tab:phaseshifters}
  \small
  \begin{tabular}{@{}lccccc@{}}
    \toprule
    \textbf{Effect} & \textbf{Loss} & $\boldsymbol{V_\pi L_\pi}$ &
    $\boldsymbol{P_\pi}$ & \textbf{Length} & \textbf{Time Response} \\
     & \textbf{(dB/mm)} & \textbf{(V$\cdot$mm)} & \textbf{(mW)} &
    \textbf{(mm)} & \textbf{(ns)} \\
    \midrule
    Electro-optical~\cite{ref32} & 0.015--0.7 & 1.8--2.8 & -- & 2--20 &
    4.4--23.3 \\
    Thermo-optical~\cite{ref33} & $<1$ & -- & 0.5--235 & 0.04--2.5 &
    $\sim(3\text{--}144)\times10^{3}$ \\
    PN junction~\cite{ref34} & 0.95--2.95 & 1.2--3.2 & -- & 2--5.5 &
    ${<}8.75\text{--}17.5$ \\
    \bottomrule
  \end{tabular}
\end{table*}

\subsection{Power Distribution Network}

Efficient and uniform optical power distribution across all antenna
elements is a critical requirement for achieving stable far-field beam
formation in optical phased arrays (OPAs). The power distribution network
divides the input light into multiple channels while minimizing excess
loss, wavelength dependence, and fabrication sensitivity.

Several splitting approaches have been demonstrated on TFLN platforms,
including Y-splitters~\cite{ref35}, multimode interference (MMI)
couplers~\cite{ref36}, and directional couplers~\cite{ref37}. Each method
offers distinct trade-offs between footprint, bandwidth, and fabrication
complexity. For the broadband operation targeted in this work
(\SIrange{1.4}{1.6}{\micro\meter}), Y-splitters were selected due to
their inherently wavelength-insensitive behavior and geometric
simplicity, which provide reliable performance without requiring complex
fabrication steps such as high-resolution inverse design.

The designed Y-splitter consists of a \SI{1}{\micro\meter}-wide input
waveguide that bifurcates into two symmetric branches through a gradual
S-shaped transition, with a total length of \SI{150}{\micro\meter}. Full
3D FDTD simulations confirmed a power imbalance below 0.3\% and excess
loss under \SI{0.2}{dB} per split across the target wavelength range. The
slight field oscillations visible near the junction arise from the
locally wider waveguide cross-section at the splitting point, which
transiently supports higher-order modes that rapidly decay beyond the
junction. Mode expansion analysis at the output of each arm confirms that
100\% of the transmitted power is carried by the fundamental TE$_{00}$
mode.

The total optical distribution network comprises nine sequential
splitting stages to feed a $25 \times 25$ antenna array. The cumulative
insertion loss introduced by these splitters is estimated to be
approximately \SI{2.7}{dB}, consistent with previously reported results
for TFLN photonic networks. This loss level ensures that, with a
\SI{10}{\milli\watt} input optical power, each antenna element receives
sufficient intensity for effective far-field emission and beam steering.

Future improvements to the distribution network may include compact
inverse-designed splitters~\cite{ref38} or low-loss multimode
interference structures to further reduce the footprint while maintaining
broadband operation. Nonetheless, the presented Y-splitter configuration
offers a robust and fabrication-tolerant solution well suited to the
proposed 3D OPA architecture.

\begin{figure*}[tp]
  \centering
  \includegraphics[width=\linewidth]{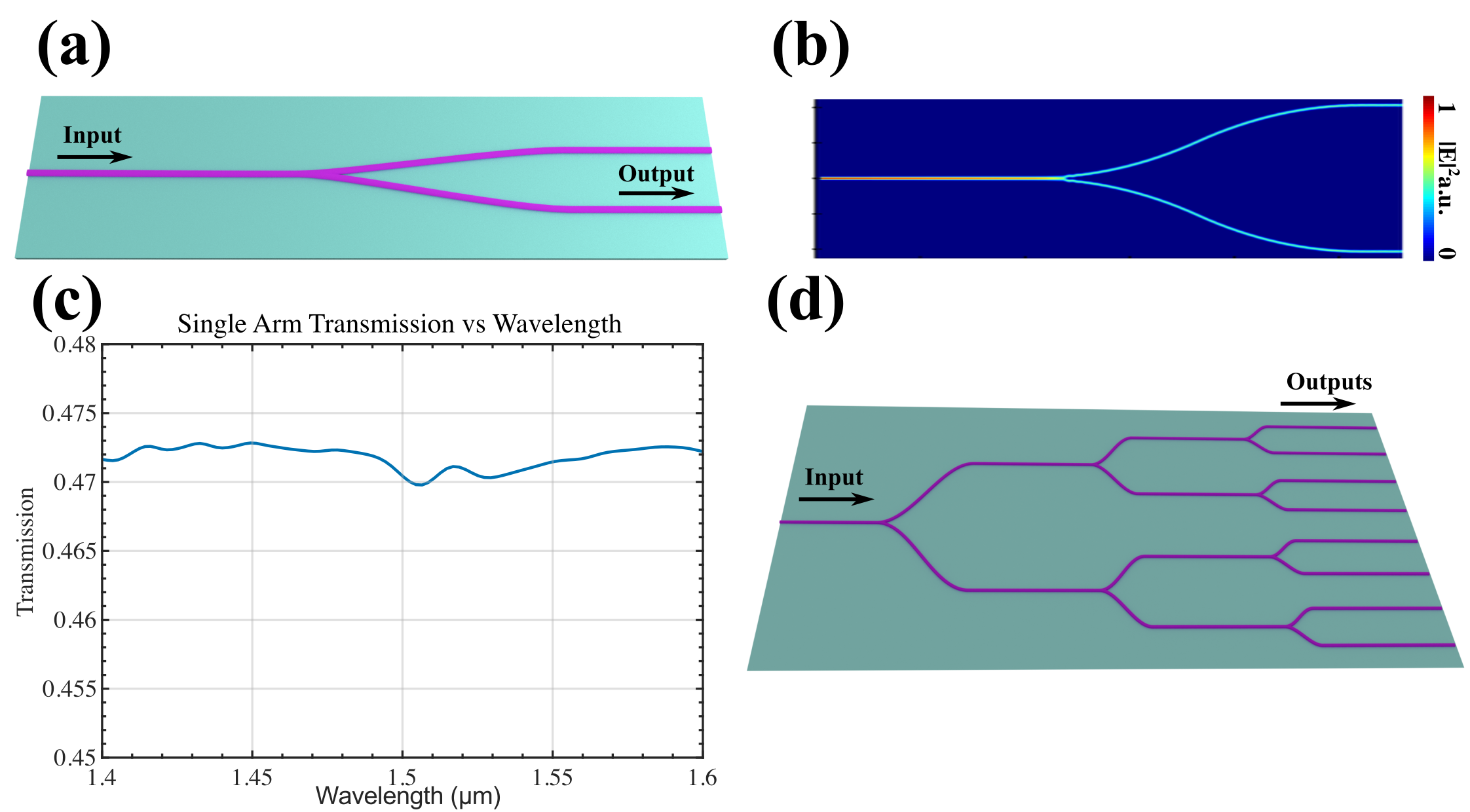}
  \caption{Y-splitter power distribution network. (a) Schematic layout.
  (b) Simulated optical field distribution. (c) Single-arm transmission
  as a function of wavelength. (d) Multi-stage splitter network used to
  feed the two-dimensional antenna array.}
  \label{fig:ysplitter}
\end{figure*}

The optimized power distribution network feeds the 3D polymer antennas
described in the following section.

\subsection{3D Antenna Architecture}

End-firing antennas are widely used in optical phased arrays (OPAs)
because they offer broadband emission and high coupling efficiency
compared to grating-based emitters, which are inherently
wavelength-sensitive. However, conventional end-fire antennas are
typically restricted to one-dimensional (1D) configurations due to their
planar geometry and edge-emission nature. To overcome this limitation, we
analyzed a three-dimensional (3D) polymer antenna fabricated directly on
a TFLN photonic platform using two-photon polymerization (2PP). This
approach enables two-dimensional (2D) antenna arrangements while
maintaining the broadband benefits of end-fire emission.

\begin{figure*}[tp]
  \centering
  \includegraphics[width=\linewidth]{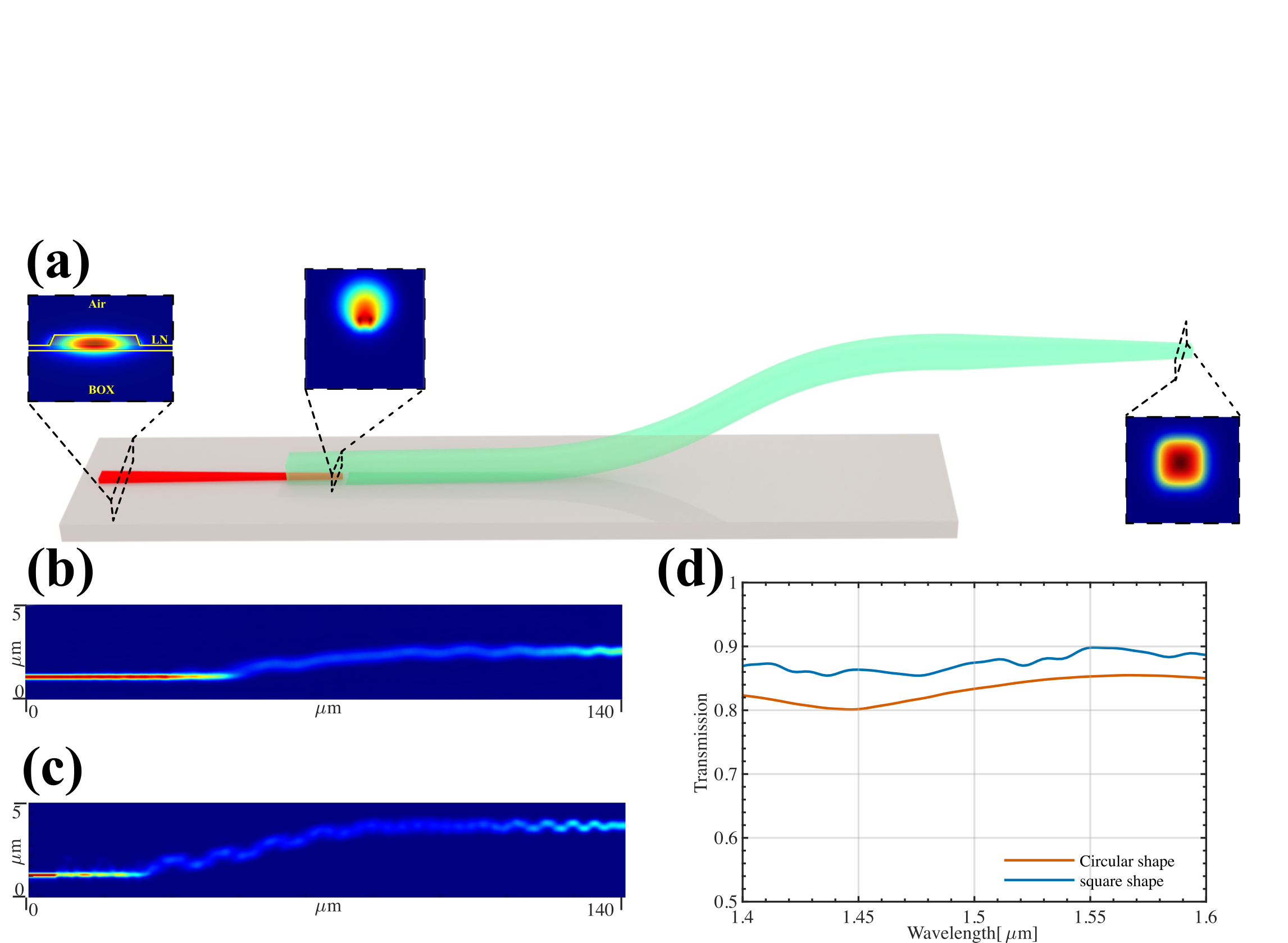}
  \caption{Three-dimensional polymer end-fire antenna design. (a)
  Geometry schematic with insets showing optical mode profiles at
  selected cross-sections. (b) Simulated mode evolution through the
  elevation section for \SI{1}{\micro\meter} square shape antenna. (c)
  Simulated mode evolution through the elevation section for
  \SI{2}{\micro\meter} square shape antenna. (d) Transmission spectrum
  over the \SIrange{1.4}{1.6}{\micro\meter} wavelength range.}
  \label{fig:antenna}
\end{figure*}

The antenna structure was designed for operation within the telecom
wavelength range of \SIrange{1.4}{1.6}{\micro\meter} and optimized for
transverse electric (TE) polarization. The polymer material, IP-n62
(refractive index $\approx 1.6$ at \SI{1.55}{\micro\meter}), offers low
optical absorption and high fabrication fidelity suitable for nanoscale
3D printing. Each antenna element comprises four adiabatically connected
sections, as illustrated in Figure~\ref{fig:antenna}(a):

\begin{enumerate}
  \item \textbf{Inverse taper coupling region}---transitions light from
  the TFLN waveguide into the polymer structure with minimal reflection
  and mode mismatch. The lithium niobate waveguide
  (\SI{1}{\micro\meter} width) narrows to a submicron tip
  ($\approx \SI{250}{\nano\meter}$) over a length of around
  \SI{30}{\micro\meter}, optimized to ensure adiabatic mode transfer from
  the TFLN waveguide to the polymer structure. The corresponding
  transition loss is estimated to be below \SI{0.5}{dB} across the
  \SIrange{1.4}{1.6}{\micro\meter} wavelength range. Critically, this
  inverse taper is positioned within the region where the polymer
  waveguide already overlaps the TFLN waveguide from above. As the TFLN
  tip narrows and its mode expands, the mode preferentially couples
  upward into the polymer cladding rather than downward into the oxide
  substrate, owing to the higher refractive index of the polymer
  ($n \approx 1.6$) compared to the oxide ($n \approx 1.44$). This ensures
  efficient and directional mode transfer into the elevating polymer
  structure.

  \item \textbf{S-curved elevation section}---lifts the optical mode
  above the substrate using a bending radius of \SI{600}{\micro\meter}
  and \ang{3} elevation angle. This geometry ensures adiabatic
  transformation while minimizing bending loss.

  \item \textbf{Adiabatic tapering section}---for the polymer structures
  with dimensions bigger than one wavelength, the polymer waveguide
  cross-section is gradually reduced to \SI{1}{\micro\meter} through an
  adiabatic 3D taper. This improves mode confinement within the waveguide
  core, promotes cleaner TE$_{00}$ mode propagation, and reduces the
  physical footprint of the emitting aperture, enabling a
  center-to-center element spacing of \SI{2}{\micro\meter} between
  neighboring antennas.

  \item \textbf{Radiating aperture}---defines the emission interface,
  controlling beam shape and far-field symmetry.
\end{enumerate}

Comprehensive 3D-FDTD simulations were used to optimize mode transfer
efficiency, elevation geometry, and aperture design. The simulated field
distribution along the propagation direction is shown in
Figure~\ref{fig:antenna}(b), demonstrating smooth mode evolution and
strong confinement throughout the elevation and radiation regions.

Quantitative results indicate a transmission efficiency of 89.5\% with
less than 5\% spectral variation across
\SIrange{1.4}{1.6}{\micro\meter}, confirming broadband operation. The
primary source of insertion loss is the mode transition from the TFLN
waveguide to the elevated polymer structure, arising from the fraction of
the optical mode that resides in the oxide substrate and therefore does
not couple into the elevating polymer waveguide. Bending loss is
negligible, as the chosen bending radius of \SI{600}{\micro\meter} far
exceeds the critical bending radius required for this waveguide geometry.
The \SI{2}{\micro\meter} square cross-section exhibited the highest
overall transmission efficiency, though with measurable multimode
interference during propagation, as quantified below. In contrast, the
\SI{2}{\micro\meter} diameter circular geometry maintained cleaner
single-mode behavior but suffered from dramatically reduced transmission
efficiency. Narrower designs ($<\SI{1.5}{\micro\meter}$) resulted in
severe coupling losses and poor performance.

To further quantify this behavior, mode-expansion analysis was performed
at the output of the adiabatic tapering section for the
\SI{2}{\micro\meter} square antenna. A fraction of the transmitted power
is carried by higher-order modes that interfere with the fundamental
TE$_{00}$ mode along the propagation length, producing the field ripple
visible in Figure~\ref{fig:antenna}(c); this ripple is most pronounced
shortly after the taper and progressively decays further along the
propagation length as the higher-order modes disperse. Extending the
taper length improves modal purity by suppressing this higher-order mode
content, at the cost of a modest reduction in total transmission
efficiency. This indicates a trade-off between total coupling efficiency
and modal purity that can be tuned via the taper length depending on
application requirements. Although extending the taper length reduces the
higher-order mode content, the resulting decrease in overall transmission
efficiency motivated the shorter optimized taper adopted in this work.

To assess the impact of mechanical support columns on modal propagation,
additional FDTD simulations were performed in which the fundamental TE
mode was excited at the input of the polymer antenna in the presence of a
support structure. A power monitor placed immediately after the support
column recorded a transmission of 99.64\%, corresponding to a propagation
loss of only 0.36\%. This confirms that the support columns do not
meaningfully perturb the propagating mode and that no optical redesign is
required (see Supplementary Fig.~S1 and Fig.~S2). The use of unoptimized
generic support structures achieving this level of modal integrity is
consistent with prior fiber-to-chip coupling demonstrations employing
similar mechanical supports~\cite{ref39}.

These results confirm that the elevated 3D antenna design effectively
bridges the on-chip TFLN waveguide and free-space domain, providing
efficient broadband coupling, low reflection, and mechanical robustness.
For mechanical stability of the elevated structures, support columns can
be straightforwardly incorporated into the 3D CAD model and fabricated in
the same 2PP printing step as the antenna itself, adding no additional
process complexity. These support strategies are standard in photonic
wire-bonding and do not introduce additional fabrication complexity.
Optical adhesive may furthermore be applied post-fabrication to
permanently fix the elevated structures without affecting their optical
performance. This optimized antenna design forms the basis for the 2D
performance analysis presented in Section~\ref{sec:results}.

\section{Results}
\label{sec:results}

Having optimized and validated the single-antenna design in
Section~2.4, we next evaluate the collective performance of the complete
optical phased array. This analysis examines how the polymer antennas
operate coherently when arranged in a two-dimensional configuration.

Using the optimized antenna geometry, electro-optic phase shifters, and
Y-splitter network described earlier, we simulated the far-field
radiation, beam steering characteristics, and overall array efficiency.
The following subsections present these results and compare the simulated
array behavior with theoretical predictions. It is worth noting that the
total OPA radiation pattern is the product of the single-element pattern
and the array factor. As shown in Figure~\ref{fig:farfield}(e), the
single antenna element produces a broad, smooth far-field pattern with no
side lobes, symmetric in both axes. The broad angular coverage of the
element pattern confirms that it does not impose any limit on the OPA
steering range---the array factor, governed by element spacing and
applied phase gradients, is the sole determinant of the achievable
steering performance.

\subsection{Array-Level Far-Field Characterization}

Based on the optimized single-element design, a two-dimensional
($N\times N$) antenna array was modeled to evaluate far-field beam
formation. Independent phase control was applied to each element,
allowing precise manipulation of the emitted field distribution. The
array simulation was performed using a rigorous two-step approach. In the
first step, the complete 3D geometry of a single polymer antenna was
simulated using full 3D FDTD, and the resulting near-field mode profile
at the antenna output was extracted using a power monitor of
\SI{3}{\micro\meter} $\times$ \SI{3}{\micro\meter}---deliberately larger
than both the waveguide cross-section and the inter-element spacing---to
capture not only the confined modal field but also any evanescent field
extending beyond the waveguide core. In the second step, this extracted
near-field profile was imported into a separate FDTD simulation
environment and replicated in a 2D array configuration, with an
independent phase assigned to each element. Crucially, because the
monitor window exceeds the inter-element spacing, any evanescent
crosstalk between neighboring elements is inherently captured and
included in the simulation.

The waveguides on the chip are separated by significantly larger
distances over most of the propagation path, and only gradually converge
to the minimum \SI{2}{\micro\meter} spacing near the antenna emission
region. Therefore, adjacent antennas experience close proximity only over
a relatively short section near the tip. To quantify the resulting
interaction, we performed additional coupling-length calculations and
full 3D FDTD simulations of representative $1\times4$ and $2\times2$
polymer antenna cells. For a constant \SI{2}{\micro\meter} spacing, the
calculated coupling length is approximately \SI{500}{\micro\meter} (see
Supplementary Fig.~S4), while the effective close-propagation region in
the actual antenna geometry is substantially shorter
($\sim\SI{140}{\micro\meter}$ maximum, with only the final tip region
reaching the minimum spacing). Consequently, only a negligible fraction
of optical power is transferred between adjacent elements. The additional
multi-element FDTD simulations confirm that the near-field amplitude and
phase distributions remain in close agreement with the two-step
single-element approach (see Supplementary Fig.~S3 and Fig.~S4). The
validation results, including multi-element FDTD simulations ($1\times4$
and $2\times2$ cells) and coupling-length analysis, are provided in
Supplementary Fig.~S3--S5. The supplementary validation includes direct
multi-element FDTD simulations ($1\times4$ and $2\times2$ cells) together
with coupling-length calculations as a function of waveguide spacing. The
structure therefore operates in the weak-coupling regime, where the
co-propagation length remains significantly smaller than the calculated
coupling length.

Far-field radiation patterns were computed for two representative element
spacings:
\begin{itemize}
  \item \SI{2}{\micro\meter} pitch, and
  \item \SI{6}{\micro\meter} pitch, representing a more widely spaced
  array.
\end{itemize}

At \SI{2}{\micro\meter} spacing, the array achieved a field of view of
\ang{24.9} $\times$ \ang{22.8} with a beamwidth of approximately
\ang{1.7}, suitable for wide-angle scanning applications. In contrast,
the \SI{6}{\micro\meter} spacing configuration produced a much narrower
\ang{0.6} $\times$ \ang{0.6} beam, reflecting the expected trade-off
between steering range and directivity.

\begin{figure}[tbp]
  \centering
  \includegraphics[width=\linewidth]{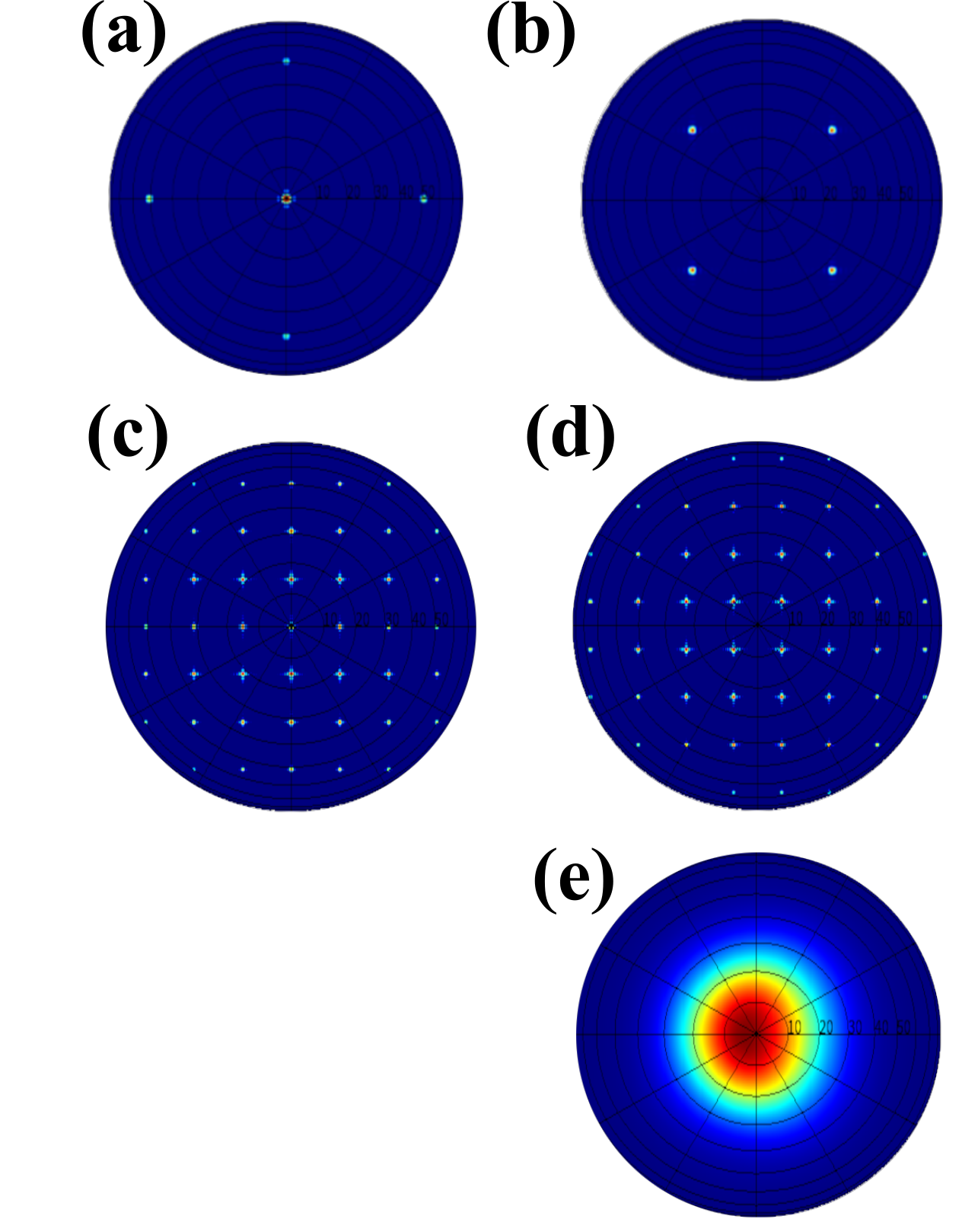}
  \caption{Simulated far-field intensity patterns of a $25 \times 25$
  optical phased array. (a) Single main lobe at \SI{2}{\micro\meter}
  element spacing with zero phase shift. (b) Steered main lobe at
  \SI{2}{\micro\meter} spacing with a \ang{180} phase shift. (c)
  Emergence of grating lobes at \SI{6}{\micro\meter} spacing with zero
  phase shift. (d) Complex interference pattern at \SI{6}{\micro\meter}
  spacing with a \ang{180} phase shift. (e) Far-field radiation pattern
  of a single antenna element, showing a broad symmetric single-lobe
  distribution confirming clean end-fire emission and demonstrating that
  the element pattern does not limit the OPA steering range.}
  \label{fig:farfield}
\end{figure}

The steering angle, $\theta$, follows the conventional array relation:
\begin{equation}
\theta = \sin^{-1}\!\left( \frac{\lambda\, \Delta\phi}{2\pi d} \right),
\end{equation}
where $\Delta\phi$ is the phase difference between adjacent elements and
$d$ is the inter-element spacing. The far-field patterns
(Figure~\ref{fig:farfield}) show strong main-lobe formation with minimal
side-lobe intensity, consistent with theoretical predictions for
uniformly excited arrays. MATLAB-based post-processing verified the
alignment between simulated and analytical results, confirming precise
phase-to-angle correspondence.

\begin{figure}[tbp]
  \centering
  \includegraphics[width=\linewidth]{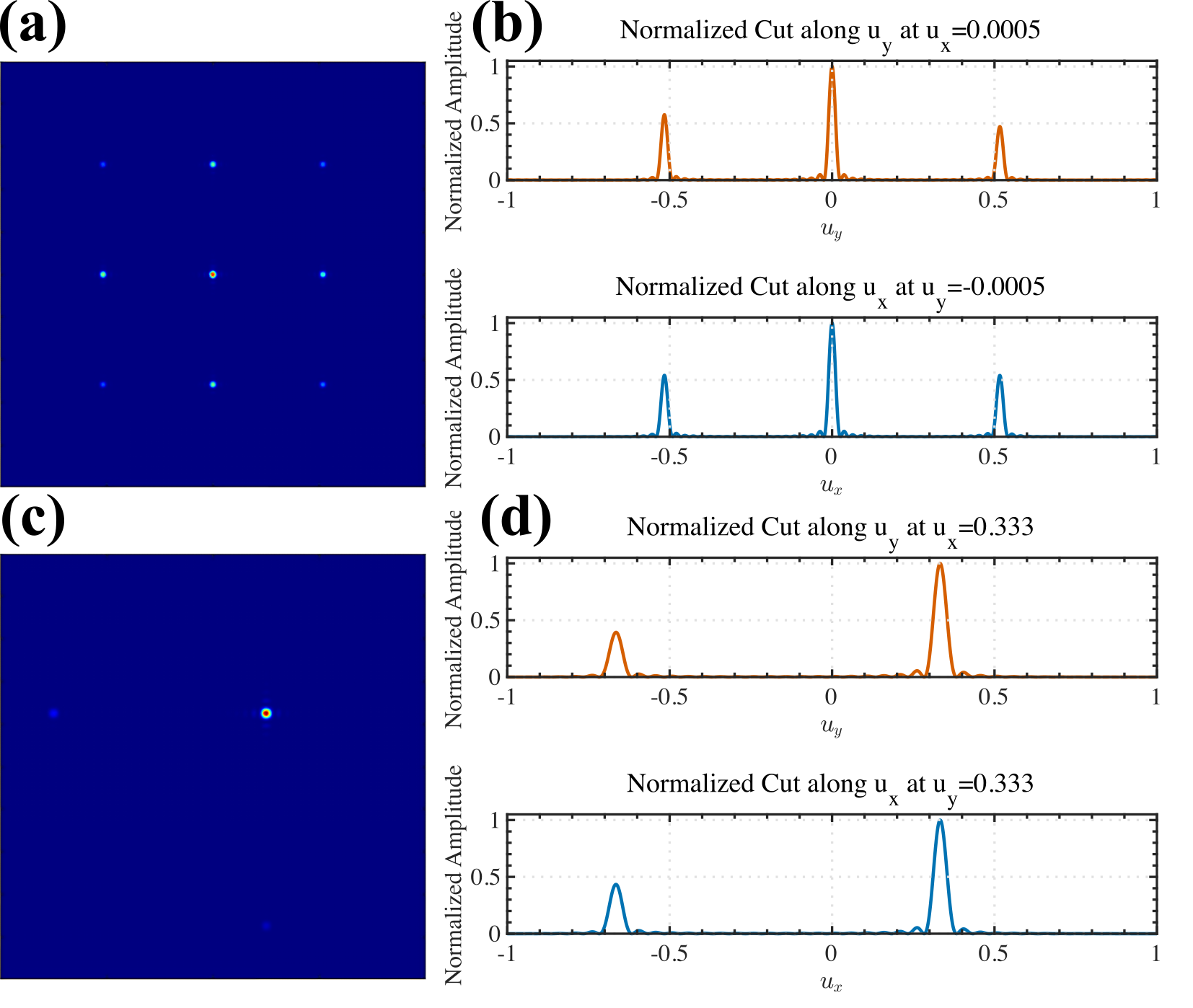}
  \caption{Beam steering and profile analysis for a $25 \times 25$ array
  with \SI{3}{\micro\meter} element spacing. (a,b) Far-field intensity
  distribution and normalized cross-sections at zero phase shift. (c,d)
  Steered beam profiles at a \ang{120} phase shift, showing displacement
  of the main lobe and changes in sidelobe structure.}
  \label{fig:steering3um}
\end{figure}

\subsection{Beam Steering and Angular Response}

Beam steering in the hybrid TFLN--polymer OPA was achieved by applying
linear phase gradients across the array through independent voltage
control of each electro-optic phase shifter. This configuration enabled
precise manipulation of the emitted optical phase front and allowed
continuous two-dimensional steering in both axes. Each EO phase shifter
can be independently driven with both a DC bias component and an AC
modulation component: the DC bias enables active correction of any static
phase errors arising from fabrication imperfections or waveguide length
variations across the array, while the AC component drives the dynamic
beam steering.

A phase shift of \ang{180} between adjacent elements---the maximum applied
in this work---corresponds to applying a voltage equal to $V_{\pi}$ to the
respective phase shifter, as defined in Section~2.2. The polymer antenna
geometries are furthermore designed to be identical across all array
elements, ensuring negligible optical path length variation; any residual
static phase offsets can be actively corrected through the DC bias
component mentioned above.

The far-field radiation characteristics of the $25 \times 25$ optical
phased array were analyzed for element spacings of \SI{2}{\micro\meter}
and \SI{6}{\micro\meter} at an operating wavelength of
\SI{1550}{\nano\meter}. Figure~\ref{fig:farfield}(a) shows that, for a
\SI{2}{\micro\meter} element spacing and uniform phase excitation, the
array produces a single dominant main lobe. Grating lobes appear at
$\pm\ang{50.70}$ for the \SI{2}{\micro\meter} spacing at zero phase.
Applying a \ang{180} phase shift results in a four-lobe interference
pattern for the same spacing, as shown in Fig.~\ref{fig:farfield}(b). In
contrast, increasing the element spacing to \SI{6}{\micro\meter} leads to
the emergence of grating lobes even under uniform phase excitation, as
illustrated in Fig.~\ref{fig:farfield}(c), while a \ang{180} phase shift
produces a more complex multi-lobe interference pattern, shown in
Fig.~\ref{fig:farfield}(d). These results highlight the strong dependence
of beam formation on both element spacing and relative phase
distribution. Figure~\ref{fig:steering3um} further illustrates the beam
steering behavior and normalized cross-sectional profiles for an
intermediate element spacing of \SI{3}{\micro\meter}, showing the main
lobe displacement and sidelobe evolution under an applied phase gradient.

Further MATLAB-based analysis of the array performance reveals the
quantitative relationship between the applied phase distribution and the
resulting steering angle, as well as the corresponding beam profiles for
various steering conditions. Controlled beam deflection through phase
manipulation is consistently observed across the steering range. It
should be noted that grating-lobe suppression is steering-angle
dependent: for the \SI{2}{\micro\meter} configuration at broadside,
grating lobes appear at $\pm\ang{50.70}$, as confirmed by the FDTD
simulations. As the beam is steered away from broadside, the main lobe
progressively shifts toward the grating lobes, reducing the available
grating-lobe-free steering range, consistent with the half-wavelength
criterion.

The main-lobe intensity decreases by less than 3\% between \ang{0} and
$\pm\ang{24.86}$, demonstrating efficient phase control and uniform
optical power distribution across the array. The measured beam-pointing
direction follows the theoretical relation
$\theta = \sin^{-1}\!\left( \frac{\lambda\, \Delta\phi}{2\pi d} \right)$
with deviations below \ang{0.3}, validating the accuracy of the
phase-to-angle conversion and confirming the precision of the EO phase
network.

To assess broadband operation, steering simulations were repeated at
three representative wavelengths: 1.45, 1.55, and \SI{1.6}{\micro\meter}.
The deflection angle varied linearly with wavelength, producing fields of
view of \ang{22.4} $\times$ \ang{20.9}, \ang{24.9} $\times$ \ang{22.8},
and \ang{25.6} $\times$ \ang{23.3}, respectively. This
wavelength-dependent behavior confirms broadband steering capability
across the telecom C-band. The polarization state and main-lobe symmetry
were preserved at all wavelengths, verifying the robustness of the 3D
polymer antenna design.

As demonstrated by the simulation methodology in Section~3.1,
inter-element crosstalk does not critically degrade array performance at
the element spacings considered, though residual crosstalk at the
smallest pitch may impose a practical limit on achievable side-lobe
suppression. The combination of strong electro-optic control and low-loss
3D antenna coupling therefore enables high-fidelity, real-time beam
steering.

Overall, the proposed OPA demonstrates broadband, low-loss, and
high-precision beam steering with a $\pm\ang{24.86}$ azimuth and
$\pm\ang{22.75}$ elevation range and excellent side-lobe suppression. The
achieved performance compares favorably with recent thin-film
lithium-niobate OPAs, which typically exhibit \ang{20}--\ang{25} steering
ranges, and with silicon-based OPAs limited to \ang{10}--\ang{30}. These
results highlight the advantages of combining TFLN's fast EO modulation
with polymer-based 3D antenna architectures to realize scalable,
high-speed beam-steering photonic systems.

\begin{figure*}[tp]
  \centering
  \includegraphics[width=\linewidth]{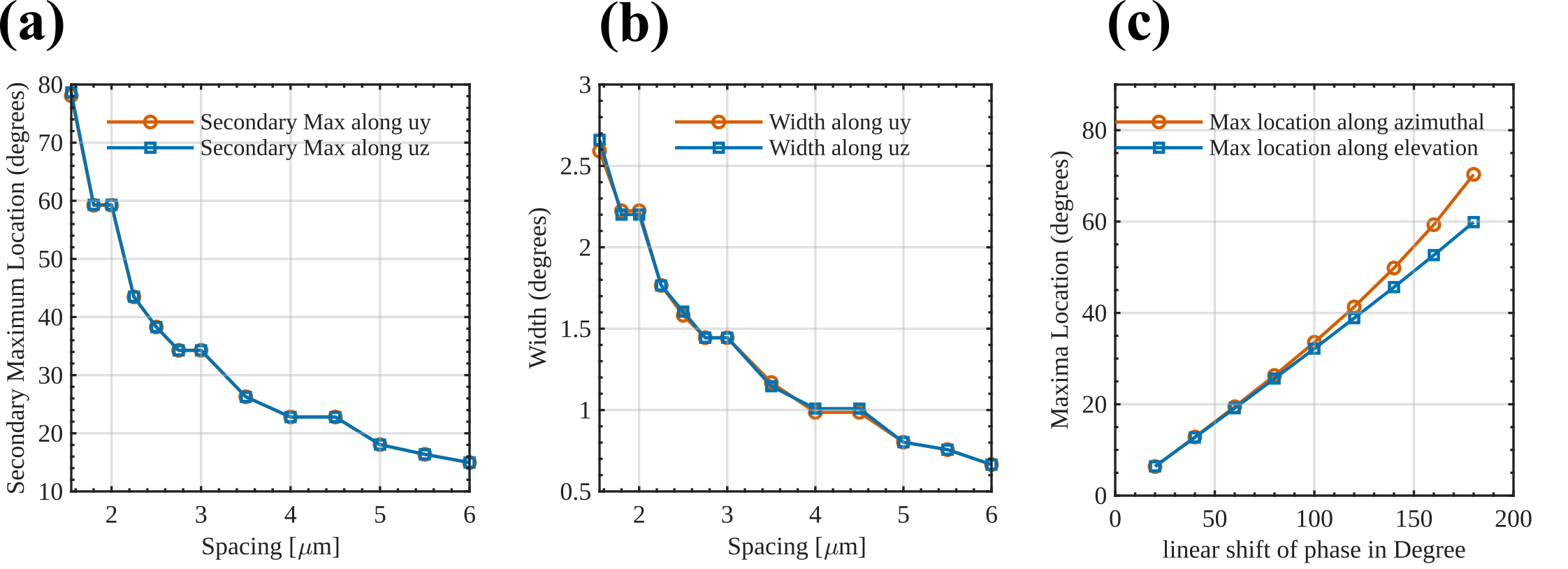}
  \caption{Performance metrics of the $25 \times 25$ phased array as a
  function of element spacing and phase shift. (a) Angular position of
  secondary maxima (grating lobes). (b) Main-lobe \SI{3}{dB} beamwidth
  versus element spacing. (c) Main-lobe steering angle as a function of
  applied linear phase gradient for \SI{2}{\micro\meter} spacing.}
  \label{fig:metrics}
\end{figure*}

\section{Discussion}

The quantitative results presented in Section~3 confirm that the proposed
three-dimensional (3D) hybrid TFLN--polymer OPA provides broadband,
high-efficiency beam steering with excellent angular precision. The
device performance arises from a careful optimization of material
platform, geometry, and fabrication parameters. The \SI{2}{\micro\meter}
square at elevation with tapering to \SI{1}{\micro\meter} polymer antenna
cross-section provides strong optical confinement and low reflection loss
while maintaining a compact inter-element pitch suitable for wide-angle
beam steering. The chosen geometry therefore represents an effective
trade-off between optical efficiency and array density.

Array-level simulations show that the OPA achieves a steering range of
$\pm\ang{24.9}$ in azimuth and $\pm\ang{22.8}$ in elevation. The field of
view (FOV) extends from \ang{22.4} $\times$ \ang{20.8} at
\SI{1.45}{\micro\meter} to \ang{25.6} $\times$ \ang{23.3} at
\SI{1.65}{\micro\meter}, confirming broadband operation. These results
compare favorably with recently reported TFLN OPAs that typically exhibit
\ang{20}--\ang{25} steering ranges, and they substantially outperform
silicon-based end-fire arrays limited to \ang{10}--\ang{30} due to
free-carrier absorption losses. The stability of the FWHM $\approx
\ang{2}$ and the small intensity drop ($<3\%$) across the steering span
demonstrate that beam quality is largely unaffected by phase modulation,
validating the efficiency of the electro-optic control network.

From a system perspective, the $25 \times 25$ array introduces an
estimated \SI{2.7}{dB} total insertion loss through the Y-splitter
distribution network, with each antenna receiving approximately
\SI{9.5}{\micro\watt} of optical power for a \SI{10}{\milli\watt} input.
The EO phase shifters leverage lithium-niobate's strong Pockels
coefficient to achieve nanosecond-scale response times, three orders of
magnitude faster than thermo-optic approaches, making this architecture
particularly suitable for high-speed LiDAR and free-space optical
communication systems.

The feasibility of the proposed 2PP antenna fabrication has been
experimentally validated through preliminary trials. A $2\times2$ polymer
antenna array was successfully fabricated on a TFLN chip using the
NanoScribe two-photon polymerization system, demonstrating the ability to
realize the elevated 3D waveguide geometry with sub-\SI{200}{\nano\meter}
feature resolution and mechanically stable structures. These results
confirm the practical viability of the fabrication approach and are
consistent with the capabilities established in the photonic wire-bonding
community. The elevated geometry not only enhances optical coupling into
free space but also opens possibilities for multilayer optical
interconnects and out-of-plane photonic integration.

Future developments of this platform could further expand its performance
envelope. Introducing non-uniform or aperiodic element spacing may
improve side-lobe suppression and extend the steering range.
Incorporating curved or concentric array geometries could enable
three-dimensional beam shaping and focal control. The demonstrated
broadband response also makes the system well suited to
wavelength-division multiplexing (WDM), allowing simultaneous multi-beam
steering at different wavelengths. Additionally, implementing polymer
optical wirebonds fabricated by the same 2PP process could replace the
current fiber-to-chip coupling scheme, simplifying packaging and
enhancing efficiency. Future work will focus on full experimental
implementation and characterization of the complete OPA system.

In summary, the proposed 3D TFLN--polymer OPA successfully combines
high-speed EO modulation with the flexibility of additive 3D fabrication,
achieving broadband, large-angle, and low-loss beam steering. This hybrid
approach establishes a practical pathway toward scalable photonic
phased-array systems for compact LiDAR, optical interconnects, and
adaptive imaging applications.

\section{Conclusion}

We have presented a novel three-dimensional (3D) end-firing OPA that
integrates elevated polymer waveguide antennas with a TFLN photonic
platform, with the 3D antenna geometry enabled through two-photon
polymerization (2PP). This hybrid architecture successfully combines the
broadband and high-efficiency characteristics of end-fire emission with
the fast electro-optic modulation capabilities of TFLN, achieving a
compact, low-loss, and scalable beam steering solution.

Through comprehensive numerical simulations, we demonstrated a
transmission efficiency of 89.5\% across the
\SIrange{1.4}{1.6}{\micro\meter} wavelength range, a field of view of
\ang{24.86} $\times$ \ang{22.75}, and beam steering up to
$\pm\ang{24.86}$ (azimuth) and $\pm\ang{22.75}$ (elevation) with
excellent side-lobe suppression and beam-pointing accuracy. The system's
electro-optic phase shifters, based on the Pockels effect, enable
nanosecond-scale response times, positioning this design as a viable
candidate for high-speed optical beam steering in LiDAR, optical
communication, and adaptive imaging systems. These results underscore the
potential of hybrid TFLN--polymer integration to bridge the gap between
integrated photonics and free-space optical systems, enabling scalable,
high-performance beam steering for future photonic technologies.

The elevated polymer antennas, realized through two-photon polymerization
(2PP) as a practical fabrication route, further extend the design
flexibility of integrated photonics, enabling true 2D array
configurations and out-of-plane coupling geometries that are challenging
to realize using conventional planar processes. The demonstrated approach
therefore establishes a scalable foundation for next-generation photonic
phased arrays that merge the advantages of integrated and free-space
optics.

Future experimental work will focus on fabrication and characterization
of the proposed device, as well as the integration of driver electronics
for real-time, high-speed beam control. Overall, the hybrid TFLN--polymer
OPA architecture provides a promising pathway toward compact, efficient,
and reconfigurable optical beam steering systems for advanced photonic
applications.

\section*{Funding}
This work was supported by the Israel Science Foundation (ISF)
[Grant No.~3291/24].

\section*{Declaration of Competing Interest}
The authors declare that they have no known competing financial interests
or personal relationships that could have appeared to influence the work
reported in this paper.

\section*{Data Availability}
Data supporting the findings of this study are available from the
corresponding author upon reasonable request.


\end{document}